\documentstyle[12pt]{article}
\begin{document}
\begin{titlepage}
\title{On the Energy-Momentum Density\\of \\Gravitational Plane Waves}
\author{ T. Dereli\footnote{E.mail: tdereli@ku.edu.tr}\\{\small Department of Physics,
Ko\c{c} University}\\{\small 34450 Sar{\i}yer,  \.{I}stanbul, Turkey} \\     \\
R. W. Tucker\footnote{E.mail: r.tucker@lancaster.ac.uk} \\
{\small Department of Physics, Lancaster University}\\ {\small
Lancaster LA1 4YB, UK}}
\date{  }

\maketitle

%\bigskip

%\bigskip

\begin{abstract}
\noindent By embedding Einstein's original formulation of GR into
a broader context we show that a dynamic covariant description of
gravitational stress-energy emerges naturally from a variational
principle. A tensor $T^G$ is constructed from a contraction of the
Bel tensor with a symmetric covariant second degree tensor field
$\Phi$ and has a form analogous to the stress-energy tensor of the
Maxwell field in an arbitrary space-time. For plane-fronted
gravitational waves helicity-2 polarised (graviton) states can be
identified carrying non-zero energy and momentum.

\bigskip


\end{abstract}
\end{titlepage}

\bigskip

\section{Introduction}

\noindent Stress and energy are invaluable concepts in Newtonian
continuum mechanics. Together with the balance laws for angular
momentum and the laws of thermodynamics they underpin the dynamics
of non-relativistic continua. They are readily assimilated into
descriptions compatible with the laws of special relativity and
the recognition that relativistic gravitation involves the
stress-energy tensor of matter on space-time was one of the key
steps in Einstein's formulation of general relativity. The
variational formulation of that theory enables one to identify the
contribution of all non-gravitational fields to the total
stress-energy tensor $T$ that enters the Einstein field equation
\begin{equation}
Ein = \frac{8\pi G}{c^4} T   \label{einstein1}
\end{equation}
where $Ein$ is the Einstein tensor and $G$ a universal coupling.
Together with the field equations for the non-gravitational fields
one has a closed system of equations for interacting gravity and
matter. In space-times with Killing symmetries, $T$ can be used to
define global conserved quantities that generalize the Newtonian
concepts above. In the absence of  such symmetry, local densities
of energy and power flux, momentum and stress  follow in frames
associated with time-like vector fields. Thus such local concepts
are covariantly defined.

From the equation above the interpretation of $T$ has no room for
the identification of local gravitational stress-energy on the
same footing as that associated with non-gravitational fields.
Thus non-trivial gravitational fields that satisfy the Einstein
field equations with $T = 0$ have no covariantly defined
mass-energy or momentum densities associated with them. This is
usually regarded as being compatible with the {\sl equivalence
principle} which relies on notions of local \lq\lq
flatness"\footnote{Regular regions of space-time are homeomorphic
to regions in $R^4$ and the tangent space at any event in the
region is isomorphic to $R^4$ as a vector space with a light-cone
structure.}
 of the space-time manifold $M$. If gravitation is associated with a non-zero
curvature tensor such arguments for the absence of local
gravitational stress-energy are not particularly persuasive.

An alternative approach is to associate {\sl gravitational} and
matter stress-energy with a class of closed 3-forms that can be
derived from the Einstein tensor. With $T = *\tau_{a} \otimes e^a$
Eqn.(\ref{einstein1}) may be written
\begin{equation}
-d\Sigma_a = \frac{8 \pi G}{c^4} \tau_a + \frac{1}{2}\left (
\omega^{b}{}_c \wedge \omega^{d}{}_a \wedge *(e_d \wedge e_b
\wedge e^c) - \omega^{d}{}_b \wedge \omega^{b}{}_c \wedge *(e_a
\wedge e_d \wedge e^c)\right )  \label{sparling}
\end{equation}
where
\begin{equation}
\Sigma_a \equiv \frac{1}{2} \omega^{b}{}_c \wedge *(e_a \wedge e_b
\wedge e^c)
\end{equation}
are the Sparling 2-forms with respect to an orthonormal co-frame
$\{e^a\}$ in which the connection forms are written
$\{\omega^{a}{}_b\}$ and $*$ denotes the Hodge map. {\footnote{
Thus if $R^a{}_{b}=\frac{1}{2}\,R^a{}_{bcd}\,e^c\wedge e^d$ then
$*\,R^a{}_{b}=\frac{1}{4}\,R^a{}_{bcd}\,\epsilon^{cd}{}_{pq}\,
e^p\wedge e^q$  in terms of the alternating symbol
$\epsilon^{cd}{}_{pq}=\eta^{ca}\eta^{db}\epsilon_{abpq}$}} A set
of pseudo-stress-energy 3-forms $\{t_a\}$ for gravity may be
chosen as
\begin{equation}
\frac{8 \pi G}{c^4} t_a = \frac{1}{2}\left ( \omega^{b}{}_c \wedge
\omega^{d}{}_a \wedge *(e_d \wedge e_b \wedge e^c) -
\omega^{d}{}_b \wedge \omega^{b}{}_c \wedge *(e_a \wedge e_d
\wedge e^c)\right )
\end{equation}
since from  Eqn.(\ref{sparling})
\begin{equation}
d(\tau_a + t_a) = 0.
\end{equation}
 However, members of this class\footnote{Members are equivalent with $t^a \sim t^a + dW^a$ for any
 set $\{W^a\}$ of 2-forms} do not
transform linearly under change of local frames and so cannot be
associated with stress-energy tensors. In this approach one must
be content with a notion of gravitational stress-energy that
depends on the coordinates used to define the local frames needed
for the construction of the components of a {\sl pseudo-tensor}.

Some years ago Bel \cite{bel}-\cite{deser1} introduced a tensor
constructed out of the space-time metric and curvature tensor that
had a number of properties in common with the stress-energy tensor
$T^{EM}$ of the source-free Maxwell field in Ricci-flat
space-times. If one writes $T^{EM} = *\tau^{EM}_{a} \otimes e^a$
in a local orthonormal coframe $\{e^a\}$ with duals
 $\{X_b\}$ ,
then
\begin{equation}
\tau^{EM}_{a} = \frac{1}{2} ( \iota_{X_a}F \wedge *F - F \wedge
\iota_{X_a}*F )
\end{equation}
 in terms of the contraction operators
$\{\iota_{X_a}\}$ and the Maxwell 2-form $F$. By contrast the Bel
tensor $B$ can be written as
\begin{equation}
B = *\tau_{a,bc} \otimes e^a \otimes e^b \otimes e^c
\end{equation}
with  associated 3-forms
\begin{equation}
\tau_{a,bc} = \frac{1}{2} ( \iota_{X_a}R_{bk} \wedge *R^{k}{}_c -
R_{bk} \wedge \iota_{X_a}*R^{k}{}_c)
\end{equation}
in terms of curvature 2-forms $\{R^{a}{}_{b}\}$.
 Since $\nabla \cdot B =0$ for the metric compatible
torsion-free connection in a Ricci-flat space-time \cite{deser2},
such a tensor has long been regarded as having something to do
with a purely gravitational stress-energy tensor. However, its
dynamical role in a theory of gravitation has remained elusive.
This is partly due to its tensorial structure and the fact that it
needs an additional factor with physical dimensions
 to relate it to an acceptable stress-energy tensor.
 In this note, we offer a means to incorporate
gravitational stress-energy into a covariant dynamical framework
in which properties of the Bel tensor enter naturally.

\section{Field Equations and Solutions}

Standard GR is based on a pseudo-Riemannian description of gravity
in which the curvature of a torsion-free connection describes the
gravitational field. A valid generalisation (made later by Brans
and Dicke) includes an additional scalar field $\phi$ as a further
gravitational degree of freedom. Here we consider instead an
additional {\it symmetric second degree covariant } tensor field
$\Phi$ and consider the gravitational action
\begin{equation}
\Lambda = \int_{M} {\cal{L}} *1
\end{equation}
where
\begin{equation}
{\cal{L}} = \frac{1}{\kappa}{\cal{R}} + R^{a}_{\,\,\,crs} R^{cbrs}
\Phi_{ab} + \frac{\lambda}{2} \nabla_{a}\Phi_{bc}
\nabla^{a}\Phi^{bc}
\end{equation}
with $R^{a}_{\,\,\,bcd}$ denoting the components of the Riemann
tensor, $\cal{R}$ the curvature scalar and $\kappa, \lambda$
constants. {\footnote{Thus by analogy with the interpretation of
the Brans-Dicke field we regard $\Phi$ as a gravitational field
that along with the metric $g$ and connection $\nabla$ determines
the geometry of spacetime in the absence of other {\it matter}
fields. If one adopts this definition the discussion in this note
is retricted to the purely gravitational sector. }} Both constants
carry physical dimensions. If the metric tensor $g$ is assigned
dimensions of length squared then $[\lambda]=L^{-2}$ and
$[\Phi^{ab}]$ has the dimensions of {\it action}.  In order to
derive field equations by variation it is convenient to write a
total action in terms of exterior forms and exterior covariant
derivatives
\begin{equation}
\Lambda^{T} = \int_{M} L = \int_{M}  \frac{1}{\kappa} R_{ab}
\wedge *(e^a \wedge e^b) + \frac{1}{2} \Phi^{ab} R_{ac} \wedge
*R^{c}_{\,\,\,b} + \frac{\lambda}{2} D\Phi_{ab} \wedge *D\Phi^{ab}
+ \frac{1}{2} F\wedge*F
\end{equation}
where the Maxwell action is included to emphasise the role of the
gravitational contribution to the total stress-energy. It proves
expedient to regard $\Lambda^T$ as a functional of $\{e^a\}$,
$\Phi$, $A$ and $\{\omega^{a}_{\,\,\,b}\}$ where $F = dA$ and
$\omega^{a}_{\,\,\,b}$ denote the connection 1-forms. So
\begin{equation}
\delta\Lambda^T = \int_{M}  \delta e^a \wedge \frac{\delta
L}{\delta e^{a}} +\delta \omega^{a}_{\,\,\,b} \wedge \frac{\delta
L}{\delta \omega^{a}_{\,\,\,b}} +\delta \Phi_{ab}\frac{\delta
L}{\delta \Phi_{ab}} +\delta A \wedge \frac{\delta L}{\delta A}
\end{equation}
where the partial variations have compact support on $M$. In this
way one determines the field equations \footnote{All variations of
$\Phi_{ab}$ induced by  orthonormal frame variations cancel due to
(\ref{phi-equation}).}
\begin{equation}
-\frac{1}{2\kappa} R^{bc}\wedge *(e_a \wedge e_b \wedge e_c) =
\tau_{a} + \lambda \tau^{\Phi}_{a} + \tau^{EM}_{a} ,
\end{equation}
\begin{equation}
\frac{1}{2\kappa} *(e_a \wedge e_b \wedge e_c) \wedge T^{c} =
\lambda ( \Phi_{ac}*D\Phi^{c}_{\,\,b} -
\Phi_{bc}*D\Phi^{c}_{\,\,a}) + \frac{1}{2}
D(\Phi_{ac}*R^{c}_{\,\,\,b}) -  \frac{1}{2}
D(\Phi_{bc}*R^{c}_{\,\,\,a}) ,\label{pop1}
\end{equation}
\begin{equation}
\lambda D*D\Phi_{ab} = \frac{1}{2} R_{ac} \wedge *R^{c}_{\,\,\,b}
, \label{phi-equation}
\end{equation}
\begin{equation}
d*F = 0
\end{equation}
where the $\Phi$ field stress-energy 3-forms
\begin{equation}
\tau^{\Phi}_{a} = \frac{1}{2} ( \iota_{X_a} D\Phi_{bc} *D\Phi^{bc}
+ D\Phi_{bc} \wedge \iota_{X_a}*D\Phi^{bc})
\end{equation}
 and $T^a = de^a + \omega^{a}_{\,\,b} \wedge e^b$ are
the torsion 2-forms. Here $\tau_{a} = \Phi^{bc} \tau_{a,bc}$ and
since the left hand side of (13) is proportional to the Einstein
tensor we identify $T^G = *\tau_a \otimes e^a$ with the
gravitational stress-energy tensor. It is worth stressing that a
variational procedure in which the connection is regarded as an
independent dynamical variable will in general give a different
system of variational field equations than one in which a
zero-torsion constraint is maintained in the variations. This is
indeed the case here due to the presence of terms that are
quadratic in components of the curvature tensor. The resulting
theory with an independent connection is simpler than that arising
from a variation with a Levi-Civita connection.

\bigskip

To substantiate the above identification  consider the  field
configuration described  in coordinates $u,v,x,y$ by the metric
\begin{equation}
g = du \otimes dv + dv \otimes du + dx \otimes dx + dy \otimes dy
+ 2 H(u,x,y) du \otimes du
\end{equation}
and
\begin{equation}
\Phi = 2 \Phi_{0} \, du \otimes du  \, , \quad A = a(u, x, y)
du,\label{19}
\end{equation}
for constant $\Phi_0$ and a torsion-free connection.{\footnote{
The sign of $\Phi_0$ is chosen below to make the energy density of
$\tau^0$, relative to a causal observer in a time-orientable
spacetime, positive.}} The symmetric tensor field $\Phi$ is null
and covariantly constant. For such a metric a useful coframe is
given by:
$$ n^1=d\,u,\quad n^2=d\,v - H\,d\,u,\quad n^3=d\,(x+iy)/\sqrt{2},\quad
n^4=d\,(x-iy)/\sqrt{2}$$  where $$g=n^1\otimes n^2 + n^2 \otimes
n^1 + n^3\otimes n^4 +n^4\otimes n^3
$$ since this has simple properties under the Hodge map associated
with $g$. It follows that the spacetime is wave-like i.e.
$R^a{}_b\wedge *\,R^b{}_c=0$ and for a covariantly constant
$\Phi$, i.e. $D\Phi=0$, it follows that (\ref{pop1}) and
(\ref{phi-equation}) are trivially satisfied.

 The
non-vanishing Maxwell stress-energy 3-forms are
\begin{equation}
\tau^{EM}_3 = \tau^{EM}_0 = \left ( \left (\frac{\partial
a}{\partial x}\right )^2 + \left ( \frac{\partial a}{\partial
y}\right )^2 \right )\, du \wedge dx \wedge dy
\end{equation}
while for the gravitational stress-energy 3-forms they are
\begin{equation}
\tau_3 = \tau_0 = \Phi_0 \left ((\Delta H)^2 - 2\,Hess(H)\right )
du \wedge dx \wedge dy
\end{equation}
with
\begin{equation}
\Delta H = \frac{\partial^2 H}{\partial x^2} + \frac{\partial^2
H}{\partial y^2}
\end{equation}
 the  Laplacian and
\begin{equation}
Hess(H) = \frac{\partial^2 H}{\partial x^2} \frac{\partial^2
H}{\partial y^2} -  \left ( \frac{\partial^2 H}{\partial x
\partial y}\right )^2
\end{equation}
 the Hessian of $H$ in  the Euclidean ($x$-$y$)-plane.

A class of plane wave solutions for $g$ and $F$ satisfying
(13)-(16) is obtained provided
\begin{equation}
\frac{1}{\kappa} \Delta H = \Phi_{0} \left ( (\Delta H)^2 -
 2 Hess(H)\right ) + \left (grad(a)\right )^2 \,  ,  \quad \Delta a =
0 \,\,.
\end{equation}
  A particular solution (with $a = 0$) is given by
\begin{equation}
H = h_{1}(u) (x^2-y^2) + 2 h_{2}(u) xy + h_{3}(u)
(x^2+y^2)\label{25}
\end{equation}
with arbitrary real functions $ h_1 , h_2 , h_3$ subject to the
condition
\begin{equation}
{h_1}^2 +  {h_2}^2 + \left ( h_3 - \frac{1}{4\kappa \Phi_0} \right
)^2 = \frac{1}{16 \kappa^2 {\Phi_0}^2} \quad .\label{26}
\end{equation}
Thus
\begin{equation}
\tau_0 = \tau_3 = 8 \Phi_0 ({h_1}^2 + {h_2}^2 +  {h_3}^2) \,\, du
\wedge dx \wedge dy .
\end{equation}
 Introducing the complex combinations $h(u) = h_1(u)
+ih_2(u)$ and $z = x + iy = re^{i \theta}$  the metric tensor
decomposes as
\begin{equation}
g = \eta +  {\cal{G}}_{+} + {\cal{G}}_{-} + {\cal{G}}_{0}
\end{equation}
 where $\eta$ is the metric of Minkowski space-time and
\begin{equation}
{\cal{G}}_{+} =  \bar{h}(u) z^2 \, du \otimes du =
\bar{{\cal{G}}}_{-} , \quad {\cal{G}}_{0} = 2 h_{3}(u)|z|^2\, du
\otimes du .
\end{equation}
 The ${\cal{G}}_{\pm}$ are null $g$-wave  helicity
eigen-tensors for linearised gravitation about $\eta +
{\cal{G}}_{0}$:
\begin{equation}
{\cal{L}}_{\frac{1}{i}\frac{\partial}{\partial
\theta}}{\cal{G}}_{\pm} = \pm 2 {\cal{G}}_{\pm}
\end{equation}
where $ {\cal {L}}_X$ denotes the Lie derivative along the vector
field $X$. It is interesting to note that the components of the
plane gravitational wave stress-energy tensor as defined above are
bounded above and below. If one writes $$\tau_0={\cal E}(u)\,
du\wedge dx\wedge dy$$ then since ({26}) is the equation of a
3-sphere and
\begin{equation}
{\cal E}(u)=\frac{2}{\kappa} h_3(u)
\end{equation}
one has $$0\le{\cal E}(u)\le \frac{1}{\kappa^2\Phi_0}.$$ Thus
causal observers will detect a bounded gravitational wave energy
flux density.

The inclusion of a polarised electromagnetic plane wave
propagating in the same direction as the gravitational wave is
straightforward. In these coordinates one takes:

$$a(u,x,y)=\alpha(u)\,x + \beta(u)\,y$$ together with (\ref{19})
and (\ref{25}). These constitute an exact solution provided
\begin{equation} {h_1}^2 +  {h_2}^2 + \left ( h_3 -
\frac{1}{4\kappa \Phi_0} \right )^2 = \left(\frac{1}{16 \kappa^2
{\Phi_0}^2}-\frac{(\alpha^2+\beta^2)}{4\Phi_0}\right) \quad .
\end{equation}
With $2P(u)=\alpha(u)-i\,\beta(u)$ one has $$ F=\frac{1}{2}
Re\left(d(P(u)\,z)\right)\wedge du$$ with
$d\left(P(u)dz\right)\wedge du$ an electromagnetic field of
helicity 1. Again one finds that ${\cal E}(u)$ is bounded but in
this case the lower bound is greater than zero and the upper bound
less than $\frac{1}{\kappa^2\Phi_0}$.

\section{Conclusion}

\noindent We have formulated an extension of GR in which the
notion of gravitational stress-energy arises naturally. It is
associated with a tensor $T^G$ constructed from a contraction of
the Bel tensor with a symmetric covariant second degree tensor
field $\Phi$ and has a form analogous to the stress-energy tensor
of the Maxwell field in an arbitrary space-time. This similarity
is particularly apparent for the case of an exact plane-fronted
gravitational wave solution in the presence of a covariantly
constant null field $\Phi$. The latter serves to endow $T^G$ with
appropriate physical dimensions. We believe that in this case
$T^G$ can be used to define gravitational stress-energy in any
causal frame and that the properties of other $\{g, \nabla,
\Phi\}$ configurations deserve further scrutiny. In particular the
static non-relativistic limit of an extension of the theory to
include matter is expected to provide a non-linear Poisson
equation for the Newtonian gravitational potential. The precise
form of such a limit will depend on how  $\Phi$ is assigned
couplings to bulk matter, an issue that must await some guidance
from experiment.

By embedding Einstein's original formulation of GR into a broader
context we have shown that a covariant description of
gravitational stress-energy emerges naturally from a variational
principle. For plane-fronted gravitational waves helicity-2
polarisation states can be identified carrying non-zero energy and
momentum. Similar states arise in GR but in the coordinates used
above such states contribute zero to the pseudo-tensor constructed
from $t_a$. The solution has been extended to include
electromagnetic plane waves constructed from helicity 1
configurations.

A number of further generalisations of the approach adopted here
may be considered including couplings of $\Phi$ to (derivatives
of) the Maxwell field and a dilaton. Such generalisations give
rise to modifications of the Maxwell equations first contemplated
by Bopp and Podolsky \cite{dereli}, Chevreton \cite{chevreton} and
more recently by Deser \cite{deser3, deser4}, Teyssandier
\cite{teyssandier1, teyssandier2}, Senovilla \cite{senovilla1} and
Benqvist et al \cite{senovilla2}. We shall discuss these
elsewhere.

\section{Acknowledgement}

TD is grateful to Lancaster University for hospitality and
acknowledges financial support from the Turkish Academy of
Sciences (TUBA). RWT is grateful to BAe Systems for partial
support. Both authors acknowledge hospitality provided by Prof. J.
Tolksdorf during the {\it Workshop on Mathematical and Physical
Aspects of Quantum Field Theories} at the H.-Fabri-Institute,
Blaubeuren, Germany  where this work was begun.


\newpage

{\small
\begin{thebibliography}{99}

\bibitem{bel} L. Bel, C.R.Acad. Sci. Paris, {\bf 247}(1958)1297

\bibitem{robinson} I. Robinson, Class. Q. Grav. {\bf 14}(1997)A331

\bibitem{deser1} S. Deser, Gen. Rel. Grav. {\bf 8}(1977)573

\bibitem{deser2} S. Deser, Class.Q. Grav. {\bf 20} (2003)L213

\bibitem{dereli} T. Dereli, G. \"{U}\c{c}oluk, Class.Q. Grav. {\bf 7}(1990)1109

\bibitem{chevreton} M. Chevreton, Nuo. Cim. {\bf 34}(1964)901

\bibitem{deser3} S. Deser, J. S. Franklin, D. Seminara, Class. Q.
Grav. {\bf 16}(1999)2815

\bibitem{deser4} S. Deser, {\it The immortal Bel-Robinson tensor},
gr-qc/9901007

\bibitem{teyssandier1} P. Teyssandier, {\it Superenergy tensors
for a massive scalar field}, gr-qc/9905080

\bibitem{teyssandier2} P. Teyssandier, Ann. d. Fond. L. de Broglie,
{\bf 26}(2001)459

\bibitem{senovilla1} J. J. M. Senovilla, Class.Q. Grav. {\bf
17}(2000)2799

\bibitem{senovilla2} G. Bergqvist, I. Eriksson, J. M. M.
Senovilla, Class. Q. Grav. {\bf 20}(2003)2663





\end{thebibliography}}
\end{document}